\documentstyle[aps,manuscript,epsfig]{revtex}
\begin{document}

\def\d{{\rm d}}
\def\reg{{\rm reg}}

\def\veck{{\pmb{k}}}
\def\vecp{{\pmb{p}}}
\def\vecv{{\pmb{v}}}

\def\vecE{{\pmb{E}}}
\def\vecF{{\pmb{F}}}
\def\vecH{{\pmb{H}}}

\title{
The Effective Lagrangian of QED with a Magnetic Charge \\
and Dyon Mass Bounds}

\author{
S. G. Kovalevich$^a$, P. Osland$^b$, Ya.\ M. Shnir$^c$, 
E. A. Tolkachev$^a$}

\address{
$^a$Institute of Physics,
Academy of Sciences of Belarus, Minsk, Belarus \\
$^b$Department of Physics, University of Bergen,
    All\'egt.~55, N-5007 Bergen, Norway \\
$^c$Department of Mathematics,
Technical University of Berlin, 10623 Berlin, Germany\thanks{
Supported by the Alexander von Humboldt
Foundation; permanent address:
Institute of Physics, 220072 Minsk, Belarus} }

\maketitle 

\thispagestyle{empty}

\bigskip

\begin{abstract}

The effective Lagrangian of QED coupled to dyons is calculated. 
The resulting generalization of the 
Euler-Heisenberg Lagrangian contains non-linear $P$ and $T$
noninvariant terms corresponding to the virtual pair creation of dyons. 
The corresponding $P$ and $T$ violating part of the matrix 
element for light-by-light scattering is considered.
This effect induces an electric dipole moment for the 
electron, of order $M^{-2}$, where $M$ is the dyon mass.
The current limit on the electric dipole moment of the 
electron yields the lower dyon mass bound $M > 1$ TeV.

\end{abstract}


\section{Introduction}
Very precise measurements achieved during the last decade
have opened up for a new approach in elementary particle physics.
According to this, evidence of new particles
can be extracted from indirect measurements of their virtual contribution
to processes at energies which are too low for direct production.
For example, the top quark mass as predicted from precision electroweak
data \cite{Ellis} agrees to within 10$\%$ with
direct experimental measurements \cite{top}.

This approach has recently been applied \cite{Rujula} for the estimation
of possible virtual monopole contributions to observables
at energies below the monopole mass. 
One-loop dyon-induced quantum corrections to the 
QED Lagrangian were discussed in \cite{Kov}. 
Taking into account the violation of parity (and time-reversal 
symmetry)
in a theory with monopoles \cite{Tom}, the emergence of an electric
dipole moment was first pointed out by Purcell
and Ramsey \cite{PurRam}.
More recently, the effect due to monopole loop contributions
has been discussed \cite{FlamMur,OslShn}.

The calculation of quantum corrections due to the virtual 
pair creation of dyons is a very difficult problem 
because the standard 
diagram technique is not valid in this case. 
The difficulty is connected both to the large value 
of the magnetic charge of the dyon and the lack of a
consistent local Lagrangian formulation of electrodynamics with two
types of charge (see e.g. \cite{StrTom} and references therein). 
So, there is no possibility to use a perturbation expansion
in a coupling constant. 
But one can apply the loop expansion which is just an expansion 
in powers of the Planck constant $\hbar$.

\section{Effective Lagrangian}
It is known (see, e.g.\ \cite{AkhiBer})
that the one-loop quantum correction to the QED Lagrangian 
can be calculated without the use of perturbation methods. 
The correction is just the change in the vacuum energy
in an external field. 
Let us review the simple case of weak constant parallel electric and
magnetic fields $\vecE$ and $\vecH$. We impose the conditions
$e |\vecE| /m^2 \ll 1$ and $e |\vecH| /m^2 \ll 1$ such that the creation
of particles is not possible. 
In this case the one-loop correction can be calculated 
by summing the one-particle modes --- the solutions of the Dirac
equation in the external electromagnetic field --- over all quantum
numbers \cite{AkhiBer}, \cite{Wentzel}. For example, if there is just
a magnetic field, $\vecH = (0, 0, H)$, 
the corresponding equation is 
\begin{equation}                                    \label{Dirac}
[i\gamma^{\mu}(\partial_{\mu} + i e A_{\mu}) - m ] \psi (x) = 0
\end{equation}
where the electromagnetic potential is
$A^{\mu} = (0, - Hy, 0, 0)$. 
The solution to this equation gives the energy
levels of an electron in a magnetic field \cite{BerLifPit,Bagrov}
\begin{equation}                                     \label{ELand}
\varepsilon _n = \sqrt{m^2 + e H (2n - 1 + s ) + k^2}
\end{equation}
where $n = 0, 1, 2 \dots $, $s = \pm 1$, and $k$ is the electron 
momentum along the field.
In this case the correction to the Lagrangian is 
\cite{AkhiBer,BerLifPit} 
\begin{eqnarray}                                      \label{DelE}
\Delta L_H = \frac{e H }{2 \pi ^2} \int \limits _0^{\infty} \d k
\left[ (m^2 + k^2) ^{1/2} 
+ 2 \sum _{n= 1} ^{\infty} (m^2 + 2eH n + k^2 )^{1/2}\right] 
\nonumber      \\
= - \frac{1}{8 \pi ^2} \int \limits _0^{\infty} \frac{\d s}{s^3} 
e^{-m^2 s} \left[
(esH)~{\rm coth}~(esH) - 1 - \frac{1}{3} e^2 s^2 H^2 \right],
\end{eqnarray}
where the terms independent of the external field
$\vecH$ are dropped and a standard renormalization 
of the electron charge has been made \cite{AkhiBer}.

If we consider simultaneously magnetic ($\vecH$) and electric ($\vecE$)
homogeneous fields, then equation (\ref{Dirac}), 
as well as its classical analogue can be separated into 
two uncoupled equations, each in two variables \cite{Bagrov}.
Indeed, in this case we can take $A^{\mu} = (Ez, - Hy, 0, 0)$ 
and the interactions of an electron with the fields
$\vecE$ and $\vecH$ are determined independently. 
For such a configuration of electromagnetic fields 
the correction to the Lagrangian is
(see \cite{AkhiBer}, p.~787)
\begin{eqnarray}                                  \label{DelE_E}
\Delta L = \frac{e H}{2 \pi ^2} \sum _{n=1}^{\infty}
\int \limits _0^{\infty} \d k \, \varepsilon _n^{(E)} (k).
\end{eqnarray}
Here $\varepsilon _n^{(E)}$ is the correction to the energy of
an electron in the combined external magnetic and electric
fields, which is in the first order proportional to $e^2 E^2$.

So, the total Lagrangian is $L = L_0 + \Delta L$, where
$L_0 = (\vecE^2 - \vecH^2)/2$~~ 
is just the Lagrangian of the free electromagnetic field
in the tree approximation, and can be written as
\begin{eqnarray}                                  \label{Ltot}
L = \left( 1 + \frac{\alpha}{3 \pi} \int \limits _{0}^{\infty}
\frac{\d s}{s} e^{-m^2 s} \right) \frac{\vecE^2 - \vecH^2}{2} 
+ \Delta L' .
\end{eqnarray}
The logarithmic divergency can be removed by the standard renormalization
of the external fields and the electron charge:
\begin{equation}                                     \label{renorm}
E_{\reg} = Z_3^{-1/2} E; \qquad H_{\reg} = Z_3^{-1/2} H; \qquad
e_{\reg} = Z_3^{1/2} e,
\end{equation}
where $Z_3^{-1} = 1 + (\alpha/3 \pi) \int \limits _{0}^{\infty}
(\d s/s) e^{-m^2 s}$ is the usual QED renormalization factor.
Thus the finite part of the correction to the Lagrangian
$\Delta L'$ can be written in terms of physical quantities as 
(see \cite{AkhiBer}, p.~790)
\begin{eqnarray}                                              
\label{DeltaL}
\Delta L'
= - \frac{1}{8 \pi ^2} \int \limits _0^{\infty} \frac{\d s}{s^3} 
e^{-m^2 s} \left[
(esE) (esH)~{\rm cot}(esE) {\rm coth}(esH) - 1
\right],
\end{eqnarray}
which in the limit $E=0$ reduces to the renormalized form
of (\ref{DelE}).

The series expansion of (\ref{DeltaL})
in terms of the parameters  $e E /m^2 \ll 1$, $e H /m^2 \ll 1$ yields
the well known Euler-Heisenberg correction \cite{Eu-Hei}:
\begin{equation}                
\label{DeltaL-Euler}
\Delta L' \approx \frac{e^4}{360 \pi ^2 m^4} 
\Bigl[ (\vecH^2 - \vecE^2)^2 +
7 (\vecH \vecE)^2 \Bigr],
\end{equation}
where $e^2=\alpha$.

Let us consider how the situation changes if we consider
the virtual pair creation of dyons in the external electromagnetic field.
Using an analogy with the classical Lorentz force on a dyon 
of velocity $\vecv$
with electric ($Q$) and magnetic ($g$) charges \cite{StrTom}
\begin{equation}
\vecF 
=  Q\vecE + g \vecH + \vecv \times (Q\vecH - g\vecE),
\end{equation}
we shall assume that the wave equation for this particle in an external
electromagnetic field can be expressed as \cite{BlagSen,Kov} 
\begin{equation}                                 \label{Dirac-dyon}
(i \gamma^{\mu} D_{\mu} - M) \psi (x) = 0,
\end{equation}
where $M$ is the dyon mass, and $iD_\mu$ 
a generalized momentum operator, with
$D_{\mu} = \partial _{\mu} + i Q A_{\mu} 
+ i g B_{\mu}$.

The potential $A_{\mu}$ and its dual $B_{\mu}$ are defined 
by $F_{\mu \nu} 
= \partial _{\mu} A_{\nu} - \partial _{\nu} A_{\mu} 
=\varepsilon _{\mu \nu \rho \sigma} \partial^{\rho} B^{\sigma}
$ where $ F_{\mu \nu} $ is the electromagnetic field strength  
tensor\footnote{ 
This definition is consistent only if $\Box 
A_{\mu} = \Box B_{\mu} = 0$, i.e., for constant electromagnetic fields 
or for free electromagnetic waves.}
and $\varepsilon_{0123}=1$. 
The potentials in the case of constant parallel electric and
magnetic fields can be expressed as
\begin{equation}                               \label{dual-pot}
A^{\mu} = (Ez, - Hy, 0, 0), \qquad B^{\mu} = (Hz, Ey, 0, 0).
\end{equation}

It is easily seen that the solution to the equation of motion for a dyon
in an external electromagnetic field can be obtained from the
solution to the equation for an electron (\ref{Dirac}) 
by the following substitution
\begin{equation}                               \label{anzatz}
e E \to Q E + g H; \qquad  e H \to Q H - g E.
\end{equation}
Using the same substitution as in Eqs.~(\ref{Ltot}) and (\ref{DeltaL}), 
we obtain the following expression for the quantum correction 
to the Lagrangian, due to the vacuum polarization caused by dyons:
\begin{eqnarray}                                 \label{Ltot-dual}
L = \left( 1 + \frac{Q^2}{12 \pi ^2} \int \limits _{0}^{\infty}
\frac{\d s}{s} e^{-M^2 s} 
- \frac{g^2}{12 \pi ^2} \int \limits _{0}^{\infty}
\frac{\d s}{s} e^{-M^2 s} \right)
\frac{\vecE^2 - \vecH^2}{2} 
+ \Delta L' ,
\end{eqnarray}
where a total derivative has been dropped.

For the renormalization of this expression we can introduce 
the renormalization factors \cite{BlagSen}
\begin{equation}                           \label{factrenorm-dual}
Z_e^{-1} = 1 + \frac{Q^2}{12 \pi} \int \limits _{0}^{\infty}
\frac{\d s}{s} e^{-M^2 s}; \quad
Z_g^{-1} = 1 - \frac{g^2}{12 \pi} \int \limits _{0}^{\infty}
\frac{\d s}{s} e^{-M^2 s},
\end{equation}
which are generalizations of the definition $Z_3$ of Eq.~(\ref{renorm}). 
In this case the fields and charges are renormalized as
\cite{BlagSen,Cal}
\begin{equation}                                               
\label{renorm-dual}
E_{\reg}^2 = Z_e^{-1}Z_g^{-1} E^2; \qquad H_{\reg}^2 
= Z_e^{-1}Z_g^{-1} H^2; \qquad
e_{\reg}^2 = Z_e Z_g e^2;\qquad g_{\reg}^2 = Z_e^{-1} Z_g^{-1} g^2.
\end{equation}
This relation (\ref{renorm-dual}) means that the vacuum of 
electrically charged particles shields the external electromagnetic field 
but the contribution from magnetically charged particles antishields it.
This agrees with the results of \cite{Giebl} and \cite{TTS}.

Considering now the case of weak electromagnetic fields,
the finite part of the Lagrangian $\Delta L'$, can, by analogy with
(\ref{DeltaL-Euler}), be written as
\begin{eqnarray}                           \label{DeltaL-dyon}
\Delta L'
& = & 
\frac{1}{360 \pi ^2 M^4} 
\bigl\{ [ (Q^2 - g^2)^2  + 7 Q^2g^2] (\vecH^2 - \vecE^2)^2 
+ [16 Q^2 g^2 + 7 (Q^2 - g^2)^2](\vecH \vecE)^2
\nonumber  \\
& &\hspace*{20mm}
+ 6 Qg (Q^2 - g^2)(\vecH \vecE)(\vecH^2 - \vecE^2) \bigr\}.
\end{eqnarray}

The expressions (\ref{DeltaL-Euler}) and (\ref{DeltaL-dyon}) 
describe nonlinear corrections to the Maxwell equations 
which correspond to photon-photon interactions.
The principal difference between the formula
(\ref{DeltaL-dyon}) and the standard Euler-Heisenberg effective Lagrangian
consists in the appearance of $P$ and $T$ non-invariant terms 
proportional to $(\vecH \vecE)(\vecH^2 - \vecE^2)$. 
It should however be noted that this term is invariant under
charge conjugation $C$, since then {\it both} $Q$ and $g$ would
change sign.

If we consider separately the virtual creation of dyon pairs,
then because of invariance of the model under a dual
transformation  (see, e.g. \cite{StrTom}) the physics is determined
not by the values $Q$ and $g$ separately, but by the effective charge
$\sqrt{Q^2 + g^2}$. 
In the same way the operations of $P$ and $T$ inversions are modified.
However we will consider simultaneously the contributions from vacuum
polarization by electron-positron and dyon pairs.
In this case it is not possible to reformulate the theory in terms of 
just one effective charge by means of a dual transformation. 
Moreover the Dirac charge quantization condition
connects just the electric charge of the electron 
and the magnetic charge of
a dyon: $eg = n/2$ whereas the electric charge $Q$ is not quantized.

It is widely believed, based both on 
experimental bounds and theoretical predictions \cite{Godd}
that the dyon mass would be large, $M \gg m$, where $m$ is the electron
mass.
Thus, in the one-loop approximation the first non-linear correction 
to the QED Lagrangian from summing the contributions
(\ref{DeltaL-Euler}) and (\ref{DeltaL-dyon}) can be written as
\begin{equation}                            \label{DeltaL-dyon-e}
\Delta L' \approx
\frac{e^4}{360 \pi ^2 m^4} \bigl[
(\vecH^2 - \vecE^2)^2 +  7 (\vecH \vecE)^2 \bigr]
+ \frac{Qg (Q^2 - g^2)}{60 \pi ^2 M^4} 
(\vecH \vecE) (\vecH^2 - \vecE^2),
\end{equation}
where the $P$ and $T$ invariant terms corresponding to vacuum polarization
by dyons have been dropped because they are suppressed by factors
$M^{-4}$. 
Thus, their contribution to the effective Lagrangian will be
of the same order as that of the ordinary QED multiloop amplitudes 
which we neglect.

\section{Photon-photon scattering}
Expression (\ref{DeltaL-dyon-e}) yields the matrix element for low-energy
photon-photon scattering. In order to determine it,
we substitute into (\ref{DeltaL-dyon-e}) the expansion
\begin{equation}
F_{\mu \nu} (x) = \frac{i}{(2\pi)^4} \int \d^4 q
\left(q_{\mu} A_{\nu} - q_{\nu} A_{\mu} \right) e^{iqx} .
\end{equation}
Corresponding to the second term of (\ref{DeltaL-dyon-e}) we find
\begin{eqnarray}                            \label{L-repr}
& &\frac{Qg (Q^2 - g^2)}{480 \pi ^2 M^4}
\int \d^4x~ \varepsilon_{\mu \nu \rho \sigma} F^{\mu \nu} F^{\rho \sigma}
F_{\alpha \beta} F^{\alpha \beta}
\nonumber\\
&=&\frac{1}{(2 \pi)^{12}}
\int \d^4 q_1 \d^4 q_2 \d^4 q_3 \d^4 q_4 \delta (q_1 + q_2 + q_3 + q_4)
A_{\mu}(q_1) A_{\nu}(q_2) A_{\rho}(q_3) A_{\sigma}(q_4) \,
\widetilde M^{\mu \nu \rho \sigma},
\end{eqnarray}
where
\begin{equation}
                                   \label{M_tilde}
\widetilde M^{\mu \nu \rho \sigma}
=\widetilde M^{\mu \nu \rho \sigma}(q_1, q_2, q_3, q_4)
=\frac{Qg(Q^2 - g^2)}{60 \pi^2 M^4}
\varepsilon_{\alpha \beta}^{\phantom{\alpha \beta} \mu \nu}
q_1^\alpha q_2^\beta
\left[q_4^\rho q_3^\sigma -g^{\rho\sigma}(q_3 q_4) \right] \, .
\end{equation}
Symmetrizing this pseudotensor one obtains the $P$ and $T$ violating part
of the matrix element for light-by-light scattering.
With all momenta flowing inwards, $k_1 + k_2 + k_3 + k_4=0$,
the matrix element takes the form
\begin{eqnarray}                                   \label{Matrix}
M_{\mu \nu \rho \sigma}'
&=&{\textstyle\frac{1}{6}} \bigl[
 \widetilde M_{\mu \nu \rho \sigma}(k_1, k_2, k_3, k_4)
+\widetilde M_{\mu \rho \nu \sigma}(k_1, k_3, k_2, k_4)
+\widetilde M_{\mu \sigma \nu \rho}(k_1, k_4, k_2, k_3)
\nonumber\\
& & 
+\widetilde M_{\nu \rho \mu \sigma}(k_2, k_3, k_1, k_4)
+\widetilde M_{\nu \sigma \mu \rho}(k_2, k_4, k_1, k_3)
+\widetilde M_{\rho \sigma \mu \nu}(k_3, k_4, k_1, k_2)
\bigr]
\nonumber\\
&=& \frac{Qg(Q^2 - g^2)}{60 \pi^2 M^4} \Bigl(
\varepsilon_{\alpha \beta }^{\phantom{\alpha \beta} \mu \nu}
k_1^\alpha k_2^\beta k_3^\sigma k_4^\rho
+ \varepsilon_{\alpha \beta }^{\phantom{\alpha \beta} \mu \rho}
k_1^\alpha k_2^\sigma k_3^\beta k_4^\nu
+ \varepsilon_{\alpha \beta }^{\phantom{\alpha \beta} \mu \sigma}
k_1^\alpha k_2^\rho k_3^\nu k_4^\beta
\nonumber\\
&&+ \varepsilon_{\alpha \beta }^{\phantom{\alpha \beta} \nu \rho}
k_1^\sigma k_2^\alpha k_3^\beta k_4^\mu
+ \varepsilon_{\alpha \beta }^{\phantom{\alpha \beta} \nu \sigma}
k_1^\rho k_2^\alpha k_3^\mu k_4^\beta
+ \varepsilon_{\alpha \beta }^{\phantom{\alpha \beta} \rho \sigma}
k_1^\nu k_2^\mu k_3^\alpha k_4^\beta \\
&&- \varepsilon_{\alpha \beta }^{\phantom{\alpha \beta} \mu \nu}
g^{\rho \sigma} (k_3 k_4) k_1^\alpha k_2^\beta
- \varepsilon_{\alpha \beta }^{\phantom{\alpha \beta} \mu \rho}
g^{\nu \sigma} (k_2 k_4) k_1^\alpha k_3^\beta
- \varepsilon_{\alpha \beta }^{\phantom{\alpha \beta} \mu \sigma}
g^{\rho \nu} (k_2 k_3) k_1^\alpha k_4^\beta
\nonumber\\
&&- \varepsilon_{\alpha \beta}^{\phantom{\alpha \beta} \nu \rho}
g^{ \mu \sigma} (k_1  k_4) k_2^\alpha k_3^\beta
- \varepsilon_{\alpha \beta }^{\phantom{\alpha \beta} \nu \sigma}
g^{ \mu \rho} (k_1 k_3)  k_2^\alpha  k_4^\beta
- \varepsilon_{\alpha \beta}^{\phantom{\alpha \beta}\rho \sigma}
g^{\mu \nu} (k_1 k_2) k_3^\alpha k_4^\beta \Bigr).
\nonumber
\end{eqnarray}
Since the interaction contains an $\varepsilon$ tensor,
the coupling between two of the
photons is different from that involving the other two,
and the familiar pairwise equivalence of the six terms does not hold.
The matrix element satisfies gauge invariance (with respect
to any of the four photons),
\begin{equation}
                                       \label{gauge-inv}
k_1^\mu\, M'_{\mu\nu\rho\sigma}(k_1,k_2,k_3,k_4)=0, \qquad \mbox{etc.}
\end{equation}
We note that the above contribution to the matrix element is proportional
to the fourth power of the inverse dyon mass,
$M'_{\mu\nu\rho\sigma}\propto M^{-4}$.
However, this result is only valid at low energies,
where the photon momenta are small compared to $M$, being obtained
from an effective, non-renormalizable theory.

Thus, as a result of interference between two one-loop diagrams
corresponding to loops with dyons and those with simply electrically
charged particles there is an asymmetry between
the processes of photon splitting and photon coalescence \cite{Kov}.
The physical effect of this asymmetry will depend on the photon
spectrum and the directions of the photon momenta with respect
to the magnetic field.  In particular, the asymmetry vanishes
when these are perpendicular, i.e. for $\cos\theta=0$.
Furthermore, the asymmetry is linear in the product of the dyon charges,
and proportional to the fourth power of the electron to dyon mass ratio.

\section{Electric dipole moment}
The contribution of this matrix element (\ref{Matrix}) breaks the $P$
and $T$ invariance of ordinary electrodynamics.
Thus, among the sixth-order radiative corrections
to the  electron-photon vertex there are terms
containing this photon-photon scattering subdiagram with a dyon loop
contribution (see Fig.~1), that induce an electric dipole moment
of the electron \cite{OslShn}\footnote{This has been noted by
I.B. Khriplovich \cite{Khrip} --- see also a recent paper by
Flambaum and Murray \cite{FlamMur}.}.

Indeed, one can  write the contribution of this diagram
to the electron-photon vertex as{\footnote{Of course,
there are more diagrams.}}:
\begin{eqnarray}           \label{vertex}
{\Lambda}_{\mu}(p',p)
&=& \frac{e^2}{(2\pi)^8} \int \d^4 k_1  \d^4 k_3
\frac {1}{k_1^2 + i\epsilon }\, \frac {1}{k_2^2 + i\epsilon }\,
\frac {1}{k_3^2 + i\epsilon}~
M_{\alpha \beta \gamma \mu}(k_1, k_2, k_3, k)
\nonumber\\
& &\times
\gamma^{\alpha}
\frac {\not p' + \not k_1 + m}{(p' + k_1)^2 -m^2 + i\epsilon}
\gamma^{\beta}
\frac {\not p - \not k_3 + m}{ (p - k_3)^2 - m^2 + i\epsilon }
\gamma^{\gamma}
\end{eqnarray}
where $M_{\alpha \beta \gamma \mu}(k_1, k_2, k_3, k)$
is the polarization pseudotensor representing
the dyon box diagram contribution to the photon-photon scattering amplitude,
the low-energy limit of which is given by the pseudotensor
$M'_{\alpha \beta \gamma \mu}$ of Eq.~(\ref{Matrix}).

In order to extract the electric dipole moment from the general
expression (\ref{vertex}) it is convenient, according to
the approach by {\cite{Kinoshita}} to exploit the identity
\begin{equation}           \label{gauge}
M'_{\alpha \beta \gamma \mu}(k_1, k_2, k_3, k)
= -k^\nu \frac{\partial}{\partial k^\mu}
M_{\alpha \beta \gamma \nu}(k_1, k_2, k_3, k),
\end{equation}
which can be obtained upon differentiating the gauge invariance condition
of the polarization tensor [cf.\ Eq.~(\ref{gauge-inv})] with respect to
$k^{\mu}$.

Substituting (\ref{gauge}) into (\ref{vertex}) we can write
the $ee\gamma$ matrix element as
\begin{eqnarray}          \label{vertex-1}
M_{ee\gamma}(p',p,k)
&=& e^\mu(k) \bar u(p') \Lambda_\mu(p',p) u(p) \nonumber \\
&=& e^\mu(k) k^{\nu} {\bar u} (p') \Lambda_{\mu \nu}(p',p) u(p)
\end{eqnarray}
where $e^{\mu}(k)$ is the photon polarization vector and
\begin{eqnarray}          \label{vertex-tens}
{\Lambda}_{\mu \nu}(p',p)
&=&-\frac{e^2}{(2\pi)^8}
\int \d^4 k_1  \d^4 k_3 ~\frac {1}{k_1^2 + i\epsilon}
~\frac {1}{k_2^2 + i\epsilon} ~\frac {1}{k_3^2 + i\epsilon}
~\frac{\partial}{\partial k^{\mu}}
M_{\alpha \beta \gamma \nu}(k_1, k_2, k_3, k)
\nonumber\\
& &\times
\gamma^{\alpha}
\frac {\not p' + \not k_1 + m}{( p' +  k_1)^2 -m^2 + i\epsilon}
\gamma^{\beta}
\frac {\not p - \not k_3 + m}{(p -  k_3)^2 - m^2 + i\epsilon}
\gamma ^{\gamma}.
\end{eqnarray}
Since the matrix element (\ref{vertex-1}) is already proportional
to the external photon momentum $k$, one can put $k=0$
in $\Lambda_{\mu \nu}$
after differentiation to obtain the static electric dipole moment.

Then, following \cite{Kinoshita}, we note that due to Lorentz covariance
of $\Lambda_{\mu \nu}$, it can be written in the form
\begin{equation}          \label{rel-repr}
{\Lambda}_{\mu \nu}(p',p)
= \left(\tilde A g_{\mu \nu}  + \tilde B \sigma_{\mu \nu}
+ \tilde C P_{\mu}\gamma_{\nu} + \tilde D P_{\nu}\gamma_{\mu}
+ \tilde E P_{\mu}P_{\nu}\right)\gamma_5
+\ldots
\end{equation}
where we have omitted terms that do not violate parity,
as well as those proportional to $k_{\mu}$, and where
$\sigma_{\mu \nu}
= (\gamma_{\mu} \gamma_{\nu} - \gamma_{\nu} \gamma_{\mu})/2$,
and
$P_{\mu} = p_{\mu} + p_{\mu}'$.

Substituting this expression into the matrix element
$M_{ee\gamma}(p',p,k)$ of Eq.~(\ref{vertex-1}) one can see that
there are two contributions to the $P$ violating part,
arising from the $\tilde B$ and $\tilde C$ terms.
In order to project out the dipole moment from (\ref{vertex-1}),
one has to compare Eq.~(\ref{rel-repr}) with the phenomenological
expression for the electric dipole moment $d_e$ \cite{Khrip-book}:
\begin{equation}         \label{mom-phen}
M_{ee\gamma}(p',p,k)=e^\mu(k) k^{\nu}
{\bar u} (p') \frac{d_e}{2m} \, \gamma _5 \, \sigma_{\mu \nu} \, u(p),
\end{equation}
In the non-relativistic limit it corresponds to the interaction 
Hamiltonian $-(d_e/2m)\vec\sigma\vec E$.
Thus, multiplying (\ref{rel-repr}) by $\sigma_{\mu\nu} \gamma _5$
and taking the trace we have:
\begin{equation}       \label{dip-mom}
d_e = -\frac{m}{24} {\rm Tr}
\left[\sigma_{\mu \nu}\gamma_5 \Lambda^{\mu \nu} \right].
\end{equation}

In order to provide an estimate of the induced electric dipole moment
we need to estimate $\Lambda^{\mu\nu}$.
The first task is to evaluate the polarization pseudotensor
$M_{\alpha \beta \gamma \mu}$ corresponding to the virtual dyon
one-loop subdiagram.
If we were to substitute for $M_{\alpha \beta \gamma \mu}$
the low-energy form $M'_{\alpha \beta \gamma \mu}$ of
Eq.~(\ref{Matrix}) into Eq.~(\ref{vertex}), we would obtain
a quadratically divergent integral.

On the other hand, straightforward application
of the Feynman rules in QED with magnetic charge
(see, e.g., \cite{BlagSen}) would give for the
photon-by-photon scattering subdiagram in Fig.~1\footnote{It
should be noted that the expression (\ref{Matrix}) contains
contributions from such loop diagrams with all possible combinations
of either three or one magnetic-coupling vertex $\Gamma_{\rho}$.},
\begin{eqnarray}                   \label{int-q}
M_{\alpha \beta \gamma \mu}(k_1, k_2, k_3, k)
&=& \frac{Q g^3}{2\pi^4} \int \d^4 q ~{\rm Tr}~\biggl(
\Gamma_{\alpha} \frac{1}{\not q + \not k_1 - M}
\Gamma_{\beta} \frac{1}{\not q - \not k_3 - \not k - M}
\nonumber\\
& & \times
\Gamma_{\gamma} \frac{1}{\not q - \not k - M}
\gamma_{\mu} \frac{1}{\not q - M} \biggr).
\end{eqnarray}
Here $\Gamma_\alpha$ represents the magnetic coupling of the photon
to the dyon, which we take according to ref.~\cite{BlagSen}
to be
\begin{equation}               \label{g-vert}
\Gamma_{\mu} = - i\varepsilon_{\mu \nu \rho \sigma}
\frac{ \gamma ^{\nu} k^{\rho} n^{\sigma}}{(n \cdot k)}.
\end{equation}
The vertex function depends on $k^{\rho}$, the photon momentum
entering the vertex, and on $n^{\sigma}$, a unit constant space-like
vector corresponding to the Dirac singularity line.
It was shown by Zwanziger \cite{Zwanz-1}
that although the matrix element depends on $n$,
the cross section as well as other physical quantities are $n$ independent.

Calculations using this technique are very complicated and can only
be done in a few simple situations \cite{BlagSen}, for example,
in the case of the  charge-monopole scattering problem \cite{Rabl}.
We will here avoid this approach.

While the integration over $q$ in Eq.~(\ref{int-q}) is logarithmically
divergent (the magnetic couplings in (\ref{int-q}) are dimensionless),
after renormalization the sum of such contributions must in the
low-energy limit reduce to the form given in Eq.~(\ref{Matrix}).
We also note that the substitution of (\ref{int-q})
into Eq.~(\ref{vertex-tens}) yields a convergent integral.
Thus, the following method for evaluating $\Lambda_{\mu\nu}$
suggests itself. We divide the region of integration into
two domains: (i) the momenta $k_1$ and $k_3$ are small compared
to $M$, and (ii) the momenta are of order $M$ (or larger).

In the first region, the form (\ref{Matrix}) can be used,
but since the integral is quadratically divergent, the integral
will be proportional to $M^2$.
Together with the over-all factor $M^{-4}$ this will give
a contribution $\propto M^{-2}$.
For large values of the photon momenta, the other form,
Eq.~(\ref{int-q}) can be used. This gives a convergent
integral, and dimensional arguments determine the scale
to be $M^{-2}$.
It means that
\begin{equation}                  \label{estim}
\left|\Lambda_{\mu \nu}\right|
\sim \frac{ e^2 Qg(Q^2 - g^2)}{(4\pi^2)^3 M^2} .
\end{equation}
The numerical coefficient has been estimated as $1/(4\pi^2)^3$,
one factor $1/4\pi^2$ from each loop, and the
$1/24$ of Eq.~(\ref{dip-mom}) is assumed cancelled by
a combinatorial factor from the number of diagrams involved.
This is of course a very rough assessment.

Now we can estimate the order of magnitude of the electron
dipole moment generated by virtual dyons.
It is obvious from Eqs.~(\ref{dip-mom}) and (\ref{estim}),
that in order of magnitude one can write
\begin{equation}            \label{dip-mom-est}
d_e \sim \frac{e^2 Qg(Q^2 - g^2)}{(4\pi^2)^3 }\frac{m}{M^2} .
\end{equation}

This estimate can be used to obtain a new bound on the dyon mass.
Indeed, recent experimental progress in the search for an
electron electric dipole moment \cite{BernSuzuki} gives a rather strict
upper limit: $d_e < 9\cdot 10^{-28} e $ cm.
If we suppose that $Q \sim e$, then from (\ref{dip-mom-est}) one can obtain
$M \geq 2\cdot 10^6~ m \approx 10^{3}$ GeV. This estimate shows
that the dyon mass belongs at least to the electroweak scale.

The above estimate coincides with the bound obtained by De R\'ujula
\cite{Rujula} for monopoles, from an analysis of LEP data, but it
is weaker than the result given in \cite{FlamMur},
where the limit $M \geq 10^{5}$ GeV was obtained.
The authors of ref.~\cite{FlamMur} used the hypothesis that a radial
magnetic field could be induced due to virtual dyon pairs.
In order to estimate the effect, they used the well-known
formula for the Ueling correction to the electrostatic potential, 
simply replacing the electron charge and mass with those of the monopole.
But the Ueling term is just a correction to the scalar Coulomb potential
due to vacuum polarization and cannot itself be considered as a source
of a radial magnetic field.
Indeed, there is only one second order term in the effective Lagrangian
that can violate the $P$ and $T$ invariance of the theory, namely
$\Delta L' \propto {\vecE}{\vecH}$.
But in the framework of QED there is no reason to consider such a correction
because it is just a total derivative.
The reference to the $\theta$-term,
used in \cite{FlamMur} to estimate the electric charge of the dyon,
is only relevant in the context of a non-trivial topology
(e.g., in the 't Hooft-Polyakov monopole model) where their limit
applies. In this case there are arguments in favour of stronger
limits on the monopole (dyon) mass (see, e.g., \cite{Godd}).

One should note that the dyon loop diagram considered above
can also contribute to the neutron electric dipole moment.
The experimental value $d_n < 1.1 \cdot  10^{-25} e $ cm \cite{Altarev}
will in the naive quark model with $m \approx 10$~MeV allow us
to obtain an estimate of the dyon lower mass bound which is similar 
to the one obtained for the electron.

\bigskip

{\bf Acknowledgements}
\medskip

We acknowledge numerous useful conversations with A. Gazizov, 
V. Kiselev and G. Calucci. 
One of us (Ya.~S.) is very indebted, for fruitful discussions,
to Prof.\ I. B. Khriplovich, to whom belongs the idea of
the above described mechanism of an electric dipole
moment generated in QED with a magnetic charge.

Ya.~S. also acknowledges support by the Alexander von Humboldt Foundation,
and, in the first stage of this work, by the 
Fundamental Research Foundation of Belarus, grant No F-094.
This research (P. O.) has been supported by 
the Research Council of Norway.

\newpage
\begin{figure}[htb]
\begin{center}
\setlength{\unitlength}{1cm}
\begin{picture}(10,6.5)
\put(-6.5,-19.0)
{\mbox{\epsfysize=30.0cm\epsffile{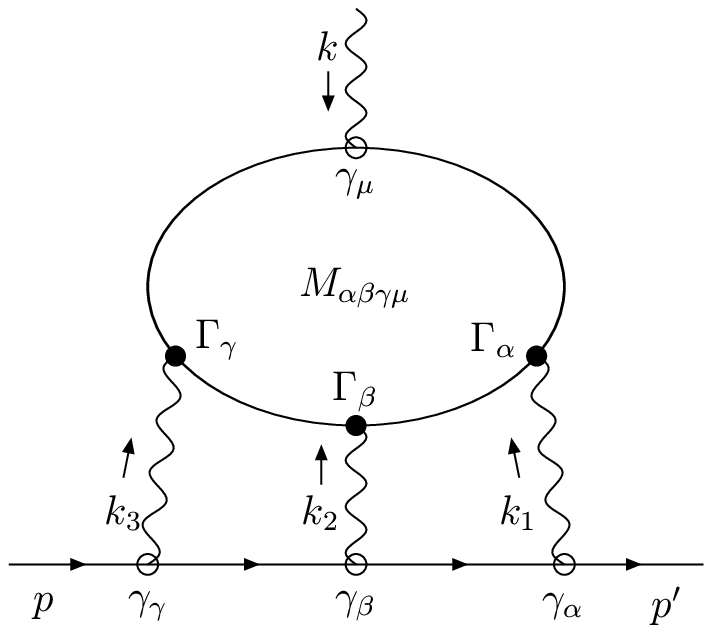}}}
\end{picture}
%
%
\caption{
Typical three-loop vertex diagram.
The closed line represents a dyon loop.}
\end{center}
\end{figure}
   
\end{document}